\documentclass[aps,showpacs,superscriptaddress,preprintnumbers]{revtex4}
\usepackage{amsfonts}
\usepackage{amssymb}
\usepackage{amsmath}
\usepackage{graphicx}
\usepackage{epsfig}

\input{tcilatex}

\begin{document}

\title{Quantum dynamics of spatial decoherence of two atoms in a ring cavity}
\author{Li~Zheng}
\affiliation{Information Science and Engineering College, Dalian Polytechnic University,
Dalian 116034, China}
\affiliation{Advanced Science Institute, RIKEN, Wako-shi, Saitama 351-0198, Japan}
\author{Chui-Ping Yang}
\affiliation{Advanced Science Institute, RIKEN, Wako-shi, Saitama 351-0198, Japan}
\affiliation{Department of Physics, University of Michigan, Ann Arbor, MI 48109, USA.}
\affiliation{Department of Physics, Hangzhou Normal University, Hangzhou, Zhejiang
310036, China}
\author{Franco Nori}
\affiliation{Advanced Science Institute, RIKEN, Wako-shi, Saitama 351-0198, Japan}
\affiliation{Department of Physics, University of Michigan, Ann Arbor, MI 48109, USA.}
\date{Nov. 3, 2010}

\begin{abstract}
We study the spatial decoherence dynamics for the relative position of two
atoms in a single-mode ring cavity. We find that the spatial decoherence of
the two atoms depends strongly on their relative position. Taking into
account the spatial degrees of freedom, we investigate the entanglement
dynamics of the internal states of the two atoms. It is shown that the
entanglement decays to almost zero in a finite time, and the disentanglement
time depends on the width of the wave packets describing the atomic spatial
distribution.
\end{abstract}

\pacs{03.65.Yz, 03.65.Ud, 42.50.Pq}
\maketitle

\section{Introduction}

Superposition and its many-particle version---entanglement---are two basic
features of quantum physics, distinguishing the quantum world from the
classical world. Because of its intriguing properties, quantum entanglement
has attracted considerable attention as an important resource for quantum
information processing \cite{ficek6,ficek7}. However, quantum coherence can
be destroyed due to the physical system interacting with the environment,
which has been recognized as a main obstacle to realizing quantum
information processing. Hence a better understanding of the mechanisms of
quantum decoherence is not only crucial for the understanding of the
quantum-classical transition (see, e.g., \cite{nori1,nori2,nori3}), but also
essential for the implementation of quantum information processing.

In the last few years, theoretical studies in this context have involved a
variety of systems (see, e.g., Refs. \cite%
{nori8,my8,my9,nori4,nori5,nori6,nori7}). Moreover, experiments have also
been done to demonstrate the dynamic process of decoherence as well as the
collapse and revival of the quantum coherence (see, e.g., Refs. \cite%
{my10,my11,my12}). In recent years, several physical systems have been
studied to learn more about environment-induced decoherence (see, e.g.,\
Refs. \cite{my15,my17}).

The influence of atomic spatial motion on quantum dynamics has been
considered in different contexts (see, e.g., Refs. \cite{wang xg,youl}). For
a system of two cold atoms placed in a noisy vacuum field, the back-action
of emitted photons on the wave packet evolution about the relative position
of the two cold atoms was discussed in Ref.  \cite{my}. It was shown that
the photon recoil resulting from the atomic spontaneous emission can induce
the localization of the relative position of the two atoms, through the
entanglement between the spatial motion of individual atoms and their
emitted photons.

In contrast with previous works, here we consider the case where a pair of
atoms simultaneously interact with a single-mode ring cavity. In fact, this
kind of system is a typical system in which the cavity field acts as a data
bus inducing two qubits to be entangled \cite{zheng,nori9}. The system is
efficiently used in many schemes for proposing how to realize quantum logic
gates and teleportation in cavity QED \cite{zheng8}. In this paper, we will
study the spatial decoherence of the atomic relative position and the
disentanglement of the internal states of these two atoms induced by the
back-action of the photons emitted from these atoms. We find that the
spatial decoherence of the two atoms depends strongly on their relative
position. Our results show that the entanglement of the internal states
decays to almost zero in a finite time, and the disentanglement time is
determined by the width of the atomic wave packets.

This paper is organized as follows. In Sec. II we present an effective
Hamiltonian of the total system and then give the solution for the
eigenequation of the effective Hamiltomian. In Sec. III we discuss the time
evolution of the density operator of the total system and obtain the reduced
density matrix in the relative-coordinate picture. Furthermore, a
decoherence factor for arbitrary cavity-field and general atomic spatial
states is introduced. In Sec. IV, under the assumption that the two atoms'
relative position is initially in a superposition state of two Gaussian wave
packets, we demonstrate the spatial decoherence when the cavity field is
initially in a coherent state. In Sec. V we study the disentanglement
dynamics of the two atoms' internal states. Finally, a concluding summary is
given in Sec. VI.

\section{Model}

\begin{figure}[b]
\includegraphics[bb=190 364 430 544,width=8cm,clip]{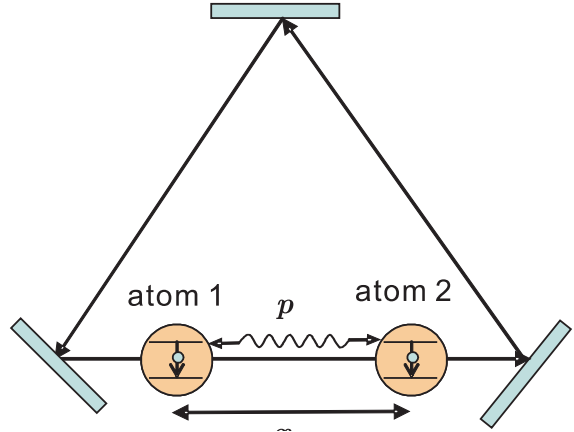}
\caption{(Color online) A schematic diagram of the system considered in this
paper. A single-mode ring cavity contains two atoms 1 and 2. Here, $p$ is
the relative momentum of the two atoms and $x$ is the relative position of
the two atoms.}
\label{figmodel}
\end{figure}

Our system consists of two identical two-level atoms that interact with a
single-mode cavity field (Fig. \ref{figmodel}). The two atoms are denoted
here as atom $1$ and atom $2.$ The mass of each atom is denoted by $m_{0}$
and their atomic transition frequency is $\omega _{0}$. Assume that the two
atoms are spatially separated and located at the positions $\hat{x}_{1}$ and
$\hat{x}_{2}$ respectively, and the corresponding momenta for atom $1$ and
atom $2$ are $\hat{p}_{1}$ and $\hat{p}_{2}$. Then the momentum for the
center of mass (c.m.) of the two atoms is $\hat{P}=\hat{p}_{1}+\hat{p}_{2}$,
and the relative momentum of the two atoms is $\hat{p}=(\hat{p}_{1}-\hat{p}%
_{2})/2$. The c.m. position and the relative position are $\hat{X}=\left(
\hat{x}_{1}+\hat{x}_{2}\right) /2$ and $\hat{x}=\hat{x}_{1}-\hat{x}_{2},$
respectively.

Under the rotating-wave approximation, the Hamiltonian of this system reads
as
\begin{equation}
\hat{H}=\frac{\hat{p}_{1}^{2}}{2m_{0}}+\frac{\hat{p}_{2}^{2}}{2m_{0}}+\frac{1%
}{2}\hbar \omega _{0}\left( \hat{\sigma}_{z}^{(1)}+\hat{\sigma}%
_{z}^{(2)}\right) +\hbar \omega \hat{a}^{\dagger }\hat{a}+\hbar g\left[ \hat{%
a}(\hat{\sigma}_{+}^{(1)}e^{ikx_{1}}+\hat{\sigma}_{+}^{(2)}e^{ikx_{2}})+%
\text{H.c.}\right] ,  \label{Hamiton}
\end{equation}%
where $\hat{\sigma}_{z}^{(i)}=\left\vert e_{i}\right\rangle \left\langle
e_{i}\right\vert -\left\vert g_{i}\right\rangle \left\langle
g_{i}\right\vert ,$ $\hat{\sigma}_{+}^{(i)}=\left\vert e_{i}\right\rangle
\left\langle g_{i}\right\vert $ and\textbf{\ }$\hat{\sigma}%
_{-}^{(i)}=\left\vert g_{i}\right\rangle \left\langle e_{i}\right\vert $ ($%
i=1,2$) are the atomic operators for the $i$-th atom with respect to the
excited state $\left\vert e\right\rangle $ and the ground state $\left\vert
g\right\rangle $ of the atoms, $\hat{a}^{\dagger }$ and $\hat{a}$ are the
creation and annihilation operators for the cavity field, $\omega $ and $k$
are the frequency and wave number of the cavity field, and $g$ is the
atom-field coupling constant. There may be a counter-propagating
running-wave mode with a wave vector $-k$ in such a ring cavity, which may
affect the atom-field dynamics and contribute to the interaction part.
However, as discussed in Ref. \cite{wang xg,sw}, we can consider that an
atom traverses only an arm of the optical ring cavity and only one cavity
mode is excited by an external laser. Thus, it is reasonable to consider a
single propagating (e.g., a clockwise-running) wave mode in a ring cavity.

We first factorize the evolution operator $\hat{U}(t)=\exp (-i\hat{H}t/\hbar
)$ into a product as
\begin{equation}
\hat{U}(t)=\hat{W}(x_{1})\hat{W}(x_{2})\hat{U}_{e}(t)\hat{W}(x_{2})^{\dagger
}\hat{W}(x_{1})^{\dagger },  \label{ut}
\end{equation}%
where $\hat{W}(x_{i})$ ($i=1,2$) is a unitary transformation defined by
\begin{equation}
\hat{W}(x_{i})=\exp \left( \frac{ikx_{i}}{2}\right) |e_{i}\rangle \langle
e_{i}|+\exp \left( \frac{-ikx_{i}}{2}\right) |g_{i}\rangle \langle g_{i}|,
\label{wx}
\end{equation}%
which concerns the coupling of the internal levels with the spatial degrees
of the atom $i$. The operator $\hat{U}_{e}(t)=\exp \left( -i\hat{H}%
_{e}t/\hbar \right) $ is easily proved to be determined by the effective
Hamiltonian $\hat{H}_{e}=\hat{H}_{0}+\hat{H}_{1}$ of the system with%
\begin{eqnarray}
\hat{H}_{0} &=&\frac{\hat{p}_{1}^{2}}{2m_{0}}+\frac{\hat{p}_{2}^{2}}{2m_{0}}+%
\frac{\hbar ^{2}k^{2}}{4m_{0}}, \\
\hat{H}_{1} &=&\hbar \sum\limits_{i=1,2}\left[ \frac{\Omega _{i}}{2}%
(\left\vert e_{i}\right\rangle \left\langle e_{i}\right\vert -\left\vert
g_{i}\right\rangle \left\langle g_{i}\right\vert )+g(\hat{a}^{\dagger
}\left\vert g_{i}\right\rangle \left\langle e_{i}\right\vert +\text{H.c.})%
\right] +\hbar \omega \hat{a}^{\dagger }\hat{a}.
\end{eqnarray}%
Here, $\Omega _{1}=\omega _{0}+p_{1}k/m_{0}$ and $\Omega _{2}=\omega
_{0}+p_{2}k/m_{0}$. In the following discussion, we consider that the atomic
transition is resonant with the cavity mode, i.e., $\omega _{0}=\omega $. In
this case, we have $\omega _{0}=ck$ ($c$ is the light velocity in vacuum).
Note that $p_{i}k/m_{0}=V_{i}k$ ($i=1,2$), where $V_{i}$ is the velocity of
the atom $i$. Thus, when the velocity of the atom $i$ is far smaller than
the light velocity in vacuum (i.e., $V_{i}<<c$), we have $p_{i}k/m_{0}<<$ $%
\omega _{0}$. On the other hand, the condition $p_{i}k/m_{0}<<g$ needs to be
met, such that the influence of the momentum-dependent energy shifts of the
atomic internal levels on the system dynamics is negligibly small. To see
the availability of this condition, let us consider a $^{87}$Rb atom with
two circular Rydberg levels $|g\rangle $ and $|e\rangle $ (corresponding to
principal quantum numbers 50 and 51). The transition frequency between $%
|g\rangle $ and $|e\rangle $ is $\sim $ 51.1 GHz \cite{add1}. Thus, we have $%
k\sim 10^{3}$ m$^{-1}$ for the case when the transition between $|g\rangle $
and $|e\rangle $ is resonant with the cavity mode. The coupling constant $g$
is on the order of 10$^{5}$ Hz \cite{add1,add2,add3}. For the laser-cooled
and optically trapped $^{87}$Rb atom, we can assume that the atom has a
measured temperature of $T\sim 180$ $\mu $K \cite{add4}. A simple
calculation shows that the atomic velocity is $V\sim 23$ c.m./s. Hence, we
have $Vk\sim 230$ Hz $<<$ 10$^{5}$ Hz, which demonstrates that the
approximation of the condition $\left\vert p_{i}k/m_{0}\right\vert <<g$
could be satisfied in practice. The analysis given here shows that the term $%
p_{i}k/m_{0}$ in $%
%TCIMACRO{\U{3a9} }%
%BeginExpansion
\Omega
%EndExpansion
_{i}$ can be neglected, leading to $%
%TCIMACRO{\U{3a9} }%
%BeginExpansion
\Omega
%EndExpansion
_{i}=\omega _{0}+p_{i}k/m_{0}$ $\approx \omega _{0}$. We note that the same
approximation was made in Ref. \cite{my}.

The total excitation number $\hat{a}^{\dagger }\hat{a}+|e_{1}\rangle \langle
e_{1}|+|e_{2}\rangle \langle e_{2}|$ is conserved during the interaction. By
resolving the eigenequation of $\hat{H}_{1},$ in the subspace spanned by
states with the total excitation number $\left( n+2\right) $, the following
eigenstates of $\hat{H}_{1}$ are obtained:
\begin{eqnarray}
\left\vert \Psi \right\rangle _{1}^{\left( n\right) } &=&\sqrt{2}%
f_{2n}\left\vert e_{1},e_{2},n\right\rangle -\sqrt{2}f_{1n}\left\vert
g_{1},g_{2},n+2\right\rangle ,  \label{pi1} \\
\left\vert \Psi \right\rangle _{2}^{\left( n\right) } &=&\frac{\sqrt{2}}{2}%
\left\vert g_{1},e_{2},n+1\right\rangle -\frac{\sqrt{2}}{2}\left\vert
e_{1},g_{2},n+1\right\rangle ,  \label{pi2} \\
\left\vert \Psi \right\rangle _{3}^{\left( n\right) } &=&f_{1n}\left\vert
e_{1},e_{2},n\right\rangle +\frac{1}{2}\left\vert
g_{1},e_{2},n+1\right\rangle  \notag \\
&&+\frac{1}{2}\left\vert e_{1},g_{2},n+1\right\rangle +f_{2n}\left\vert
g_{1},g_{2},n+2\right\rangle ,  \label{pi3} \\
\left\vert \Psi \right\rangle _{4}^{\left( n\right) } &=&-f_{1n}\left\vert
e_{1},e_{2},n\right\rangle +\frac{1}{2}\left\vert
g_{1},e_{2},n+1\right\rangle  \notag \\
&&+\frac{1}{2}\left\vert e_{1},g_{2},n+1\right\rangle -f_{2n}\left\vert
g_{1},g_{2},n+2\right\rangle ,  \label{pi4}
\end{eqnarray}%
with the eigenvalues
\begin{equation}
E_{1,2}^{\left( n\right) }=(n+1)\hbar \omega =E_{0},\qquad E_{3}^{\left(
n\right) }=E_{0}+\hbar A_{n},\qquad E_{4}^{\left( n\right) }=E_{0}-\hbar
A_{n},  \label{e4}
\end{equation}%
where $A_{n}=\sqrt{2(2n+3)}g$, $f_{1n}=\sqrt{\left( n+1\right) /\left[
2\left( 2n+3\right) \right] }$, $f_{2n}=\sqrt{\left( n+2\right) /\left[
2\left( 2n+3\right) \right] }$ and $n$ is an arbitrary non-negative integer.

\section{Decoherence factor}

In this section, we discuss the time evolution of the system, and
investigate the spatial decoherence factor for the atom-atom relative
position.

Assume that the initial density operator of the whole system is given by
\begin{equation}
\hat{\rho}\left( 0\right) =\hat{\rho}_{s}\!\left( 0\right) \;\hat{\rho}%
_{i}\!\left( 0\right) \;\hat{\rho}_{f}\!\left( 0\right) ,  \label{density 0}
\end{equation}%
where $\hat{\rho}_{s}\left( 0\right) =\left\vert \psi \left( 0\right)
\right\rangle \left\langle \psi \left( 0\right) \right\vert $ is the initial
density operator for the spatial motion of the two atoms, $\hat{\rho}%
_{i}\left( 0\right) =\left\vert \phi \left( 0\right) \right\rangle
\left\langle \phi \left( 0\right) \right\vert $ is the initial density
operator for the internal state of the two atoms, and $\hat{\rho}_{f}\left(
0\right) $ is the initial density operator for the cavity field.

Let us now assume that the state $\left\vert \psi \left( 0\right)
\right\rangle $ is expanded (in the momentum representation) as
\begin{equation}
\left\vert \psi \left( 0\right) \right\rangle =\int \int_{-\infty }^{\infty
}dp_{1}dp_{2}\;C_{p_{1}p_{2}}\left\vert p_{1},p_{2}\right\rangle ,
\label{initial p}
\end{equation}%
where $\left\vert p_{1}\right\rangle $ is the momentum eigenstate of atom $1$%
, $\left\vert p_{2}\right\rangle $ is the momentum eigenstate of atom $2,$
and $C_{p_{1}p_{2}}$ is the distribution function satisfying the
normalization condition$\int \int_{-\infty }^{\infty }dp_{1}dp_{2}\left\vert
C_{p_{1}p_{2}}\right\vert ^{2}=1.$ The state $\left\vert \phi \left(
0\right) \right\rangle $ is assumed to be a pure separable state $\left\vert
e_{1},e_{2}\right\rangle $. In terms of Fock states $\left\vert
n\right\rangle $ and $\left\vert n^{\prime }\right\rangle ,$ the density
operator $\hat{\rho}_{f}\left( 0\right) $ is, in general, written as
\begin{equation}
\hat{\rho}_{f}\left( 0\right) =\underset{n,n^{\prime }}{\sum }c_{n,n^{\prime
}}\left\vert n\right\rangle \left\langle n^{\prime }\right\vert .
\label{field initial}
\end{equation}

The time evolution of $\hat{\rho}\left( 0\right) $ can be obtained by $\hat{%
\rho}\left( t\right) =\hat{U}(t)\hat{\rho}\left( 0\right) \hat{U}^{\dagger
}(t).$ Following the above eigenstates (\ref{pi1})-(\ref{pi4}), eigenvalues (%
\ref{e4}) of $\hat{H}_{1},$ and based on Eqs. (\ref{ut}-\ref{wx}) and (\ref%
{density 0}-\ref{field initial}), we find that the density operator $\hat{%
\rho}\left( t\right) $ can be written as
\begin{equation}
\hat{\rho}\left( t\right) =\underset{n,n^{\prime }}{\sum }c_{n,n^{\prime
}}\left\vert \Psi _{n}(t)\right\rangle \left\langle \Psi _{n^{\prime
}}(t)\right\vert ,  \label{dendity of t}
\end{equation}%
where%
\begin{eqnarray}
\left\vert \Psi _{n}(t)\right\rangle  &=&\int \!\int_{-\infty }^{\infty
}dp_{1}dp_{2}\;C_{p_{1}p_{2}}\exp \left\{ -\frac{it}{2m_{0}\hbar }\left[
\left( p_{1}-\frac{\hbar k}{2}\right) ^{2}+\left( p_{2}-\frac{\hbar k}{2}%
\right) ^{2}\right] \right\}   \notag \\
&&\times \lbrack D_{1}\left( n,t\right) \left\vert p_{1},p_{2}\right\rangle
\otimes \left\vert e_{1},e_{2},n\right\rangle +D_{2}\left( n,t\right)
(\left\vert p_{1}-\hbar k,p_{2}\right\rangle \otimes \left\vert
g_{1},e_{2},n+1\right\rangle   \notag \\
&&+\left\vert p_{1},p_{2}-\hbar k\right\rangle \otimes \left\vert
e_{1},g_{2},n+1\right\rangle )+D_{3}\left( n,t\right) \left\vert p_{1}-\hbar
k,p_{2}-\hbar k\right\rangle \otimes \left\vert g_{1},g_{2},n+2\right\rangle
]  \label{persit}
\end{eqnarray}%
and%
\begin{eqnarray}
D_{1}\left( n,t\right)  &=&2f_{1n}^{2}\cos \left( A_{n}t\right) +2f_{2n}^{2},
\label{d1} \\
D_{2}\left( n,t\right)  &=&-if_{1n}\sin \left( A_{n}t\right) ,  \label{d2} \\
D_{3}\left( n,t\right)  &=&2f_{1n}f_{2n}\Big[\cos \left( A_{n}t\right) -1%
\Big].  \label{d3}
\end{eqnarray}%
We assume that the distribution function $C_{p_{1}p_{2}}$ in Eq.~(\ref%
{initial p}) satisfies $C_{p_{1}p_{2}}=C_{P}C_{p}$; i.e., the initial state $%
\left\vert \psi \left( 0\right) \right\rangle $ for the spatial motion of
the two atoms can be written as $\left\vert \psi \left( 0\right)
\right\rangle =\left\vert \mu \left( 0\right) \right\rangle \otimes
\left\vert \varphi \left( 0\right) \right\rangle $, with $\left\vert \mu
\left( 0\right) \right\rangle =\int_{-\infty }^{\infty }dP\;C_{P}\left\vert
P\right\rangle $ describing the initial c.m. state of the atoms with a
momentum distribution function $C_{P}$ corresponding to the c.m. momentum
eigenstate $\left\vert P\right\rangle $, and $\left\vert \varphi \left(
0\right) \right\rangle =\int_{-\infty }^{\infty }dp\;C_{p}\left\vert
p\right\rangle $ describing the initial relative position state of the atoms
with a distribution function $C_{p}$ corresponding to the relative momentum
eigenstate $\left\vert p\right\rangle $. Both $C_{P}$ and $C_{p}$ satisfy
the normalization condition, i.e., $\int_{-\infty }^{\infty }\left\vert
C_{P}\right\vert ^{2}dP=1$ and $\int_{-\infty }^{\infty }\left\vert
C_{p}\right\vert ^{2}dp=1.$

By tracing over the cavity field, the c.m. motion, and the internal states
of the atoms, we obtain from Eq.~(\ref{dendity of t}) the following elements
of the reduced density matrix (in a relative coordinate picture)
\begin{equation}
\rho \left( x,x^{\prime },t\right) =\varphi \left( x,t\right) \varphi ^{\ast
}\left( x^{\prime },t\right) F\left( x,x^{\prime },t\right) ,
\label{reduced density}
\end{equation}%
where $\varphi \left( x,t\right) =\int_{-\infty }^{\infty }dp\;C_{p}\;\exp
\left\{ -ip^{2}t/\left( m_{0}\hbar \right) +ipx/\hbar \right\} $ is the
free-evolution state of $\left\vert \varphi \left( 0\right) \right\rangle $
expressed in a relative coordinate picture, and $F\left( x,x^{\prime
},t\right) $ is the decoherence factor, which is given by
\begin{equation}
F\left( x,x^{\prime },t\right) =\underset{n=0}{\sum^{\infty }}c_{n,n}\left\{
D_{1}^{2}\left( n,t\right) +D_{3}^{2}\left( n,t\right) +2\left\vert
D_{2}\left( n,t\right) \right\vert ^{2}\cos \left[ k(x-x^{\prime })/2\right]
\right\} .{}  \label{decoherence factor}
\end{equation}%
The decoherence factor $F\left( x,x^{\prime },t\right) $ given here will be
used below in our analysis of the spatial decoherence of two atomic wave
packets.

\section{Spatial decoherence of two atomic wave packets}

We consider that the initial state $\left\vert \varphi \left( 0\right)
\right\rangle $ of the two atoms' relative position is a superposition of
two Gaussian wave packets centered at $a$ and $-a,$ respectively; i.e., we
have (in the $x$ representation)
\begin{equation}
\varphi \left( x,0\right) =\frac{1}{\sqrt{2\delta }}\left[ G_{+}(x)+G_{-}(x)%
\right] ,  \label{fi0}
\end{equation}%
\begin{figure}[b]
\includegraphics[bb=92 337 436 597,width=8cm,clip]{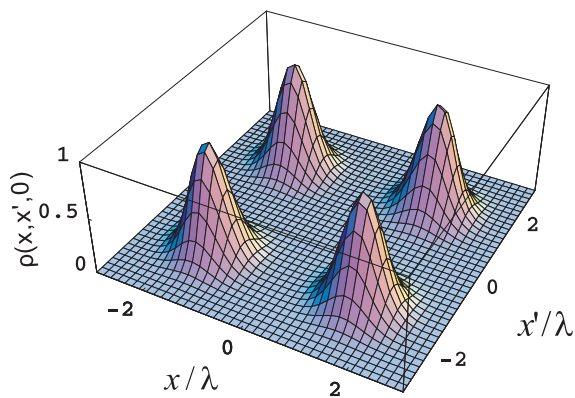}
\caption{(Color online) Plot of the density matrix $\protect\rho \left(
x,x^{\prime },0\right) $ representing the initial state for the relative
position of the two atoms. It is a superposition of two Gaussian wave
packets. The two peaks along the $x^{\prime }=-x$ diagonal direction
represent the coherence between the two wave packets.}
\label{figinitial}
\end{figure}
where $G_{\pm }(x)=\left( \sqrt{2\pi }d\right) ^{-1/2}\exp \left[ -(x\pm
a)^{2}/\left( 4d^{2}\right) \right] $ and $\delta =1+\exp \left[
-a^{2}/\left( 2d^{2}\right) \right] $. Here $\delta $ is a normalization
constant and, for simplicity, we have assumed that the two Gaussian
distributions have the same spread $d$, which is limited in $d_{\min }\leq
d\leq a.$ Here, $d_{\min }$ is much smaller than the wavelength $\lambda $
of the cavity field, but a zero spread is not permitted to avoid the atoms
indistinguishable by quantum dispersion \cite{my1}.

Assume that the time for the atoms staying in the cavity is so short that we
can assume the spatial distribution remains unchanged with respect to the
initial spatial state for the atomic free evolution. Then Eq.~(\ref{reduced
density}) can be written as a product of the initial state and the
decoherence factor%
\begin{equation}
\rho \!\left( x,x^{\prime },t\right) \simeq \varphi \!\left( x,0\right)
\varphi ^{\ast }\!\left( x^{\prime },0\right) F\!\left( x,x^{\prime
},t\right) =\rho \!\left( x,x^{\prime },0\right) F\!\left( x,x^{\prime
},t\right) ,  \label{reduced density 0}
\end{equation}%
where $\rho \left( x,x^{\prime },0\right) =\varphi \left( x,0\right) \varphi
^{\ast }\left( x^{\prime },0\right) $ describes the initial state of the
atomic relative position. The initial state is illustrated in Fig. \ref%
{figinitial}. Here, the two peaks, along the $x^{\prime }=x$ direction,
correspond to the diagonal terms of $\rho \left( x,x^{\prime },0\right) $,
while the other two peaks along the $x^{\prime }=-x$ direction correspond to
the off-diagonal terms of $\rho \left( x,x^{\prime },0\right) ,$ which
represent the coherence between the two wave packets.

Following Eq.\negthinspace \thinspace (\ref{decoherence factor}), it can be
proved that for the case of $x^{\prime }=x$, $F\left( x,x,t\right) =1,$ thus
we have $\rho \left( x,x,t\right) =\rho \left( x,x,0\right) $, i.e., the
diagonal terms of the density matrix $\rho \left( x,x^{\prime },t\right) $
remain unchanged during the time evolution. However, for the off-diagonal
terms with $x^{\prime }=-x\neq m\lambda $ ($m$ is an arbitrary nonzero
integer), it can be seen that $\left\vert F\left( x^{\prime }=-x,t\right)
\right\vert \leq 1.$ Thus we will mainly analyze the evolution of the
off-diagonal terms of the reduced density matrix below.\textit{\ }

For the cavity field initially in a coherent state $\left\vert \alpha
\right\rangle $, we have $c_{n,n}=\exp \left( -|\alpha |^{2}\right) \alpha
^{2n}/n!$, where $\alpha $ is a complex number. Thus, Eq. (\ref{decoherence
factor}) becomes
\begin{equation}
F\left( x,x^{\prime },t\right) =e^{-|\alpha |^{2}}\underset{n=0}{%
\sum^{\infty }}\frac{\alpha ^{2n}}{n!}\left\{ D_{1}^{2}\left( n,t\right)
+2\left\vert D_{2}\left( n,t\right) \right\vert ^{2}\cos \left[ \frac{%
k(x-x^{\prime })}{2}\right] +D_{3}^{2}\left( n,t\right) \right\} .{}
\end{equation}%
Along the $x^{\prime }=-x$ direction, the decoherence factor takes the
following form%
\begin{equation}
F\left( x^{\prime }=-x,t\right) =e^{-|\alpha |^{2}}\underset{n=0}{%
\sum^{\infty }}\frac{\alpha ^{2n}}{n!}\left\{ D_{1}^{2}\left( n,t\right)
+2\left\vert D_{2}\left( n,t\right) \right\vert ^{2}\cos \left( kx\right)
+D_{3}^{2}\left( n,t\right) \right\} ,{}
\end{equation}%
which is shown in Fig. \ref{figcoherent}. It can be seen that:
\begin{figure}[tph]
\includegraphics[bb=3 1 320 625,width=8cm,clip]{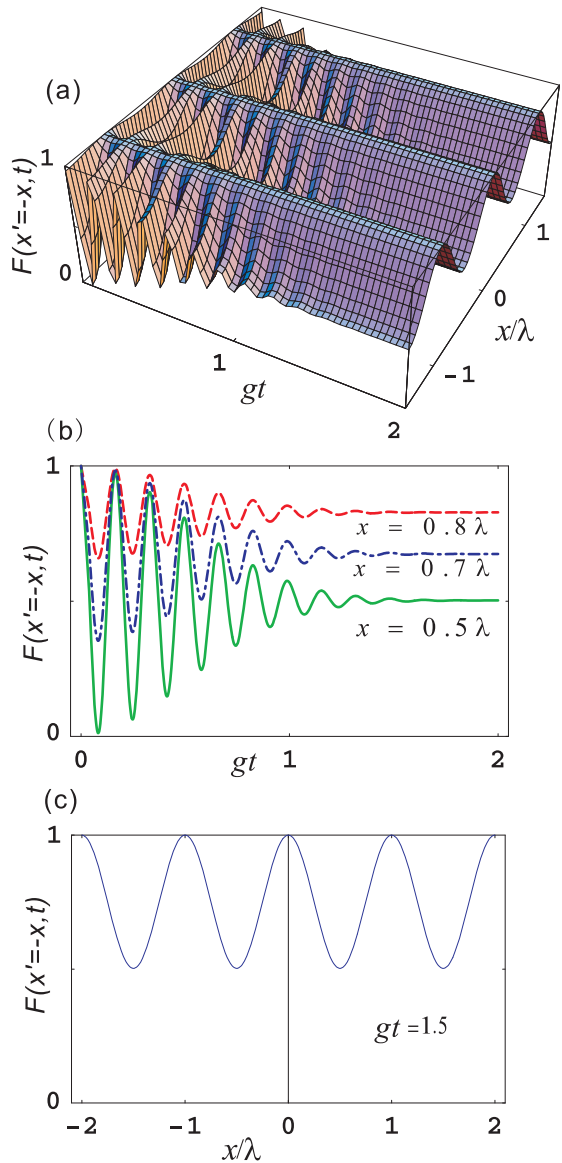}
\caption{(Color online) (a) Evolution of the decoherence factor $F(x^{\prime
}=-x,t)$ with time $t$ and relative position $x$. Decoherence factor $%
F(x^{\prime }=-x,t)$ for (b) some selected relative positions and (c) a
selected time$.$ (a), (b) and (c) were ploted for the cavity field initially
in a coherent state with $\protect\alpha =10$.\ }
\label{figcoherent}
\end{figure}

(i) for $x=m\lambda $, $F(x^{\prime }=-x,t)$ remains equal to $1$ during the
time evolution [Fig. \ref{figcoherent}(a)];

(ii) for $x\neq m\lambda $, $F(x^{\prime }=-x,t)$ decays to a constant after
a finite time and the value of the constant is determined by the value of $x$
[Fig. \ref{figcoherent}(b)].

(iii) For a given time $t$, $F(x^{\prime }=-x,t)$ oscillates with $x$ in a
cosine law [Fig. \ref{figcoherent}(c)], and for the case of $x=\left(
m+1/2\right) \lambda $, the decay of $F(x^{\prime }=-x,t)$ reaches the
maximum with $F(x^{\prime }=-x,gt\geqslant 1.5)$ $\approx $ $0.5$ [Fig. \ref%
{figcoherent}(c)].

Above, we presented an analysis on the decoherence factor, for the\ case of
the initial state $\varphi (x,0)$ described by Eq. (\ref{fi0}). According to
the above analysis, it can be concluded that:

(i) for the case $a=m\lambda $, the two peaks along the $x^{\prime }=-x$
direction remain unchanged during the time evolution. So the relative motion
of the two atoms decouples with the cavity field, and thus there is no
decoherence induced by the photon recoil.

(ii) for the case $a\neq m\lambda $, the two peaks along the $x^{\prime }=-x$
direction partially decay in a finite time and there is no revival, and the
decay of the two peaks depends on $a$ in a cosine law, with a maximum decay
at $a=\left( m+1/2\right) \lambda .$%
\begin{figure}[tbh]
\includegraphics[bb=69 178 378 394,width=8cm,clip]{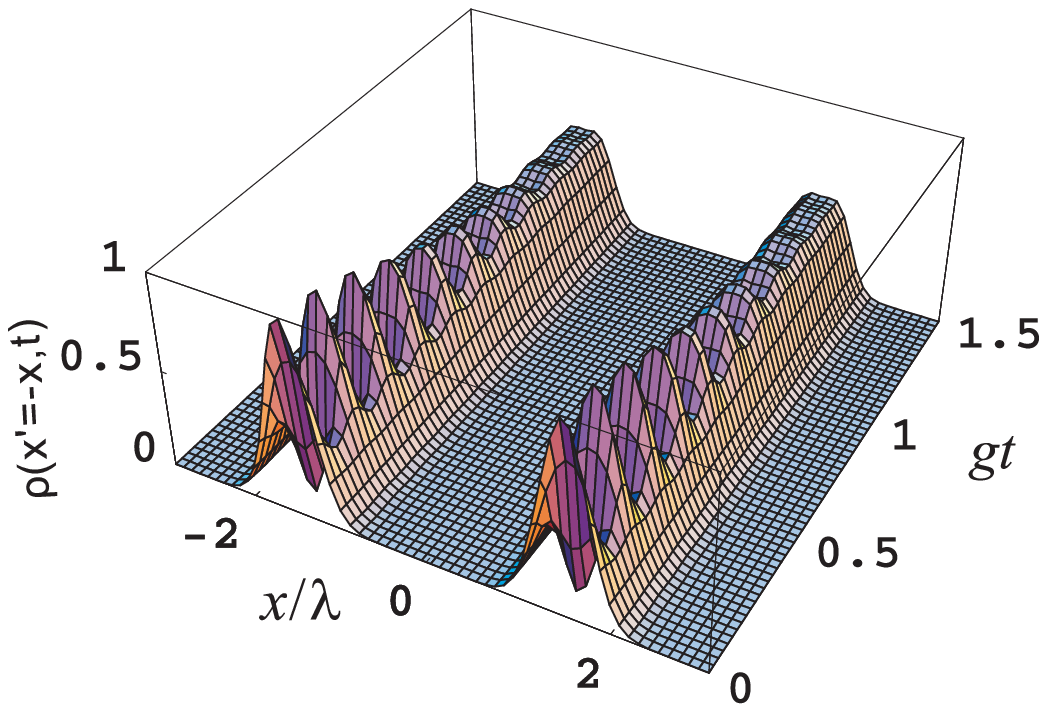}
\caption{(Color online) Density matrix $\protect\rho (x^{\prime }=-x,t)$,
which evolves from the initial state (\protect\ref{fi0}) with $a\neq m%
\protect\lambda $ for the cavity field initially in a coherent state with $%
\protect\alpha =10$.\ }
\label{figdemon}
\end{figure}
\begin{figure}[tb]
\includegraphics[bb=81 271 491 721,width=8cm,clip]{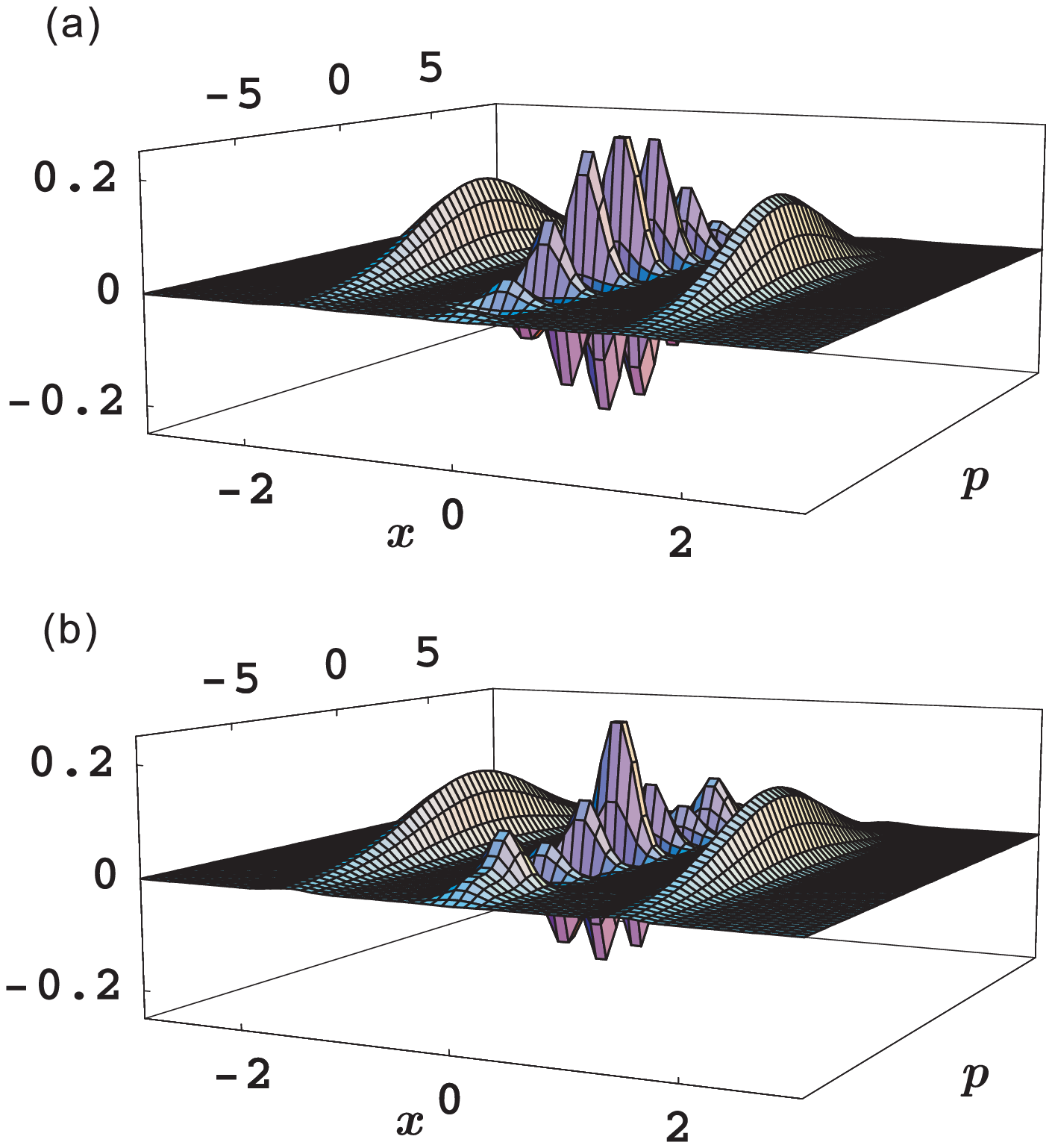}
\caption{(Color online) (a) The Wigner function for the initial density
matrix $\protect\rho \left( x,x^{\prime },0\right) $ shown in Fig. \protect
\ref{figinitial}. (b) The Wigner function for the density matrix in
Eq.\thinspace\ (\protect\ref{reduced density 0}) whose off-diagonal terms
were shown in Fig. \protect\ref{figdemon}. (a) and (b) were plotted for $%
gt=2.$ Here, $p$ is the relative momentum of the two atoms and $x$ is the
relative position of the two atoms.}
\label{figwigner}
\end{figure}

The evolution of $\rho (x^{^{\prime }}=-x,t)$ for the initial state $\varphi
(x,0)$ with $a\neq m\lambda $ is shown in Fig. \ref{figdemon}, which
demonstrates that the initial state $\varphi (x,0)$ undergoes a partial
decoherence after a finite time and a revival does not occur.\ Thus, the
coherence of the atomic relative position is strongly destroyed. The wigner
functions for the initial density matrix $\rho (x,x^{^{\prime }},0)$ shown
in Fig. \ref{figinitial} and the density matrix $\rho (x,x^{^{\prime }},t)$
given in Eq. (\ref{reduced density 0}) are shown in Fig. \ref{figwigner}.
Strong oscillations together with negative values [Fig. \ref{figwigner}(a)]
indicate quantum coherence between the two wave packets, and it turns out
that the oscillations are partially damped by decoherence [Fig. \ref%
{figwigner}(b)]. This may be induced by the entanglement between the
two-atom spatial motion and the cavity field resulting from the photonic
back-action. This entangling process inevitably destroys the coherence of
the spatial motion of the two atoms. Since the single-mode cavity field is
just an environment with a few degrees of freedom, the spatial coherence is
not completely destroyed.

\section{Disentanglement dynamics of two atoms}

To begin with, let us assume that:

(i) the initial spatial state $\left\vert \psi \left( 0\right) \right\rangle
$ of the two atoms is a separable state
\begin{equation}
\left\vert \psi \left( 0\right) \right\rangle =\left\vert \mu \right\rangle
_{1}\otimes \left\vert \mu \right\rangle _{2}.
\end{equation}%
Here, $\left\vert \mu \right\rangle _{1}$ and $\left\vert \mu \right\rangle
_{2}$ represent two Gaussian wave packets describing the spatial
distribution of the two atoms, respectively. In the coordinate
representation, $\left\vert \mu \right\rangle _{i}$ is expressed as ($i=1,2$)%
$:$%
\begin{equation}
\mu \left( x_{i},0\right) =\int_{-\infty }^{\infty }dp_{i}\;C_{p_{i}}\;\exp
\left( \frac{i}{\hbar }p_{i}x_{i}\right) =\left( \frac{1}{2\pi d^{2}}\right)
^{1/4}\exp \left[ -\frac{(x_{i}+a_{i})^{2}}{4d^{2}}\right] ,
\end{equation}%
where $a_{i}$ is the center of the Gaussian function $\mu \left(
x_{i},0\right) $ and $d$ is the width of the Gaussian function $\mu \left(
x_{i},0\right) $. \ The coefficient $C_{p_{i}}$ is given by
\begin{equation}
C_{p_{i}}=\left( \frac{2d^{2}}{\pi \hbar ^{2}}\right) ^{1/4}\exp \left( -%
\frac{d^{2}p_{i}^{2}}{\hbar ^{2}}+\frac{ia_{i}p_{i}}{\hbar }\right) .
\label{cpd entangle2}
\end{equation}%
For convenience, we assume that $a_{1}=-a_{2}=a/2$. The initial density
operator for the spatial motion of the two atoms is $\hat{\rho}_{s}\left(
0\right) =\left\vert \psi \left( 0\right) \right\rangle \left\langle \psi
\left( 0\right) \right\vert $.

(ii) the initial internal state $\left\vert \phi \left( 0\right)
\right\rangle $ of the two atoms is an entangled state, which is given by
\begin{equation}
\left\vert \phi \left( 0\right) \right\rangle =\cos \gamma \left\vert
g_{1},g_{2}\right\rangle +\sin \gamma \left\vert e_{1},e_{2}\right\rangle .
\label{entangle state}
\end{equation}%
The initial density operator for the internal state of the two atoms is $%
\hat{\rho}_{i}\left( 0\right) =\left\vert \phi \left( 0\right) \right\rangle
\left\langle \phi \left( 0\right) \right\vert .$

(iii) the cavity field is initially in a vacuum state, i.e., $\hat{\rho}%
_{f}\left( 0\right) =\left\vert 0\right\rangle \left\langle 0\right\vert $.

The initial density operator of the whole system\ is $\hat{\rho}\left(
0\right) =\hat{\rho}_{s}\!\left( 0\right) \;\hat{\rho}_{i}\!\left( 0\right)
\;\hat{\rho}_{f}\!\left( 0\right) $. At time $t$, the density operator of
the system is $\hat{\rho}\left( t\right) =\hat{U}(t)\hat{\rho}\left(
0\right) \hat{U}^{\dagger }(t)$. Here, $\hat{U}(t)$ is the unitary operator
in Eq.\thinspace (\ref{ut}). After tracing $\hat{\rho}\left( t\right) $ over
the electromagnetic field and the spatial degrees of freedom of the atoms,
we obtain the following reduced density matrix for the internal state of the
two atoms:
\begin{equation}
\hat{\rho}_{i}(t)=\left(
\begin{array}{cccc}
a & 0 & 0 & w \\
0 & b & z & 0 \\
0 & z^{\ast } & c & 0 \\
w^{\ast } & 0 & 0 & d%
\end{array}%
\right) ,  \label{reduced den en}
\end{equation}%
which is written in a basis formed by $\left\vert e_{1},e_{2}\right\rangle $%
, $\left\vert e_{1},g_{2}\right\rangle $, $\left\vert
g_{1},e_{2}\right\rangle $ and $\left\vert g_{1},g_{2}\right\rangle $. Here$%
, $
\begin{eqnarray*}
a &=&D_{1}^{2}(0,0)\sin ^{2}\gamma , \\
b &=&c=\left\vert D_{2}(0,0)\right\vert ^{2}\sin ^{2}\gamma , \\
d &=&\cos ^{2}\gamma +D_{3}^{2}(0,0)\sin ^{2}\gamma , \\
w &=&D_{1}(0,0)\cos \gamma \sin \gamma e^{-i2\omega t}e^{-s(t)}, \\
z &=&be^{-\xi (t)},
\end{eqnarray*}%
with $\xi (t)=s(t)+d^{2}k^{2}-ika$ and $s(t)=\hbar ^{2}k^{2}t^{2}/\left(
4d^{2}m_{0}^{2}\right) $. The parameter $s(t)$ is obtained after tracing $%
\hat{\rho}\left( t\right) $ over the atomic spatial degrees of freedom,
which causes the strong time dependence of the concurrence as shown below.

\begin{figure}[tbp]
\includegraphics[bb=136 329 447 532,width=8cm,clip]{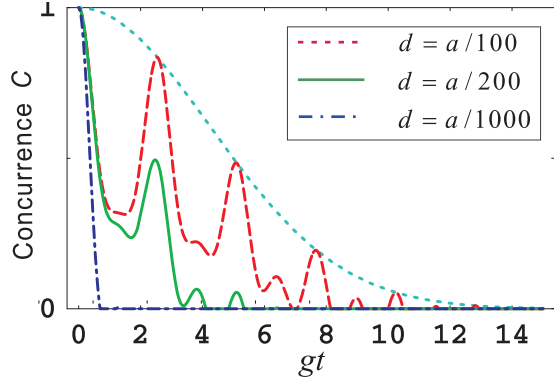}
\caption{(Color online) Time evolution of the concurrence $C$ for the case
when the initial internal state of the two atoms is an entangled state $%
\left\vert \protect\phi \left( 0\right) \right\rangle =\cos \protect\gamma %
\left\vert g_{1},g_{2}\right\rangle +\sin \protect\gamma \left\vert
e_{1},e_{2}\right\rangle $, with $\protect\gamma =\protect\pi /4.$ The
dotted line is for $e^{-s}$ with $d=a/100.$}
\label{figdisentanglement}
\end{figure}

A popular measure of entanglement for $\hat{\rho}_{i}(t)$ is given by the
concurrence \cite{concurrence}, which is defined as
\begin{equation}
C\left( \hat{\rho}_{i}\right) =\max \left\{ 0,\sqrt{\lambda _{1}}-\sqrt{%
\lambda _{2}}-\sqrt{\lambda _{3}}-\sqrt{\lambda _{4}}\right\} ,
\label{concurrence formula}
\end{equation}%
where the $\lambda _{i}$ are the eigenvalues of the non-Hermitian matrix $%
\hat{\rho}_{i}\widetilde{\hat{\rho}}_{i}$ in descending order, and $%
\widetilde{\hat{\rho}}_{i}=\left( \hat{\sigma}_{y}^{1}\otimes \hat{\sigma}%
_{y}^{2}\right) \hat{\rho}_{i}^{\ast }\left( \hat{\sigma}_{y}^{1}\otimes
\hat{\sigma}_{y}^{2}\right) .$ Here, $\hat{\rho}_{i}^{\ast }$ is the complex
conjugate of $\hat{\rho}_{i},$ and $\hat{\sigma}_{y}^{i}$ is the usual Pauli
operator for atom $i$ ($i=1,2$). The concurrence $C$ varies from $C=0$ for a
non-entangled state to $C=1$ for a maximally-entangled state. The
concurrence of the density matrix $\hat{\rho}_{i}(t)$ in Eq. (\ref{reduced
den en}) is \cite{tingyu,yuting2}%
\begin{equation}
C\left( \hat{\rho}_{i}(t)\right) =2\max \{0,\left\vert z\right\vert -\sqrt{ad%
},\left\vert w\right\vert -\sqrt{bc}\},  \label{c1}
\end{equation}%
which is shown in Fig. \ref{figdisentanglement}. Eq. (\ref{c1}) tells us
that the concurrence evolves with an non-smoothly exponential decay due to
the factor $e^{-s}$ as shown in Fig. \ref{figdisentanglement}$.$

Alternatively, the initial internal state for the two atoms could be another
type of entangled state given by
\begin{equation}
\left\vert \phi \left( 0\right) \right\rangle =\cos \gamma \left\vert
e_{1},g_{2}\right\rangle +\sin \gamma \left\vert g_{1},e_{2}\right\rangle .
\end{equation}%
By doing a calculation similar to the one above, it can be seen that for the
entanglement state $\left\vert \phi \left( 0\right) \right\rangle $
considered here, the reduced density matrix $\hat{\rho}_{i}(t)$ for the
internal state of the two atoms is
\begin{equation}
\hat{\rho}_{i}(t)=\left(
\begin{array}{cccc}
0 & 0 & 0 & 0 \\
0 & b & z & 0 \\
0 & z^{\ast } & c & 0 \\
0 & 0 & 0 & w%
\end{array}%
\right) ,  \label{partial density}
\end{equation}%
where
\begin{eqnarray*}
b &=&A_{+}^{2}(t)\cos ^{2}\gamma +A_{-}^{2}(t)\sin ^{2}\gamma \\
&&+2A_{+}(t)A_{-}(t)\cos \gamma \sin \gamma e^{-\delta }\cos \left(
ak\right) , \\
c &=&A_{-}^{2}(t)\cos ^{2}\gamma +A_{+}^{2}(t)\sin ^{2}\gamma \\
&&+2A_{+}(t)A_{-}(t)\cos \gamma \sin \gamma e^{-\delta }\cos (ak), \\
w &=&\left\vert B(t)\right\vert ^{2}+2\cos \gamma \sin \gamma e^{-\delta
}\cos \left( ak\right) , \\
z &=&e^{-s(t)}\left\{ A_{+}(t)A_{-}(t)\left[ \cos ^{2}\gamma e^{-\alpha
}+\sin ^{2}\gamma e^{-\beta }\right] \right. \\
&&\left. +\left[ A_{+}^{2}(t)+A_{-}^{2}(t)e^{-\eta }\right] \cos \gamma \sin
\gamma \right\} ,
\end{eqnarray*}%
with
\begin{eqnarray*}
A_{\pm }(t) &=&\frac{\cos (\sqrt{2}gt)\pm 1}{2}, \\
B(t) &=&-i\sin (\sqrt{2}gt)/\sqrt{2}.
\end{eqnarray*}%
Here, $\delta =d^{2}k^{2},$ $\alpha =d^{2}k^{2}+iak,\beta =d^{2}k^{2}-iak$
and $\eta =4d^{2}k^{2}+i2ak.$

\begin{figure}[tbp]
\includegraphics[bb=136 329 447 532,width=8cm,clip]{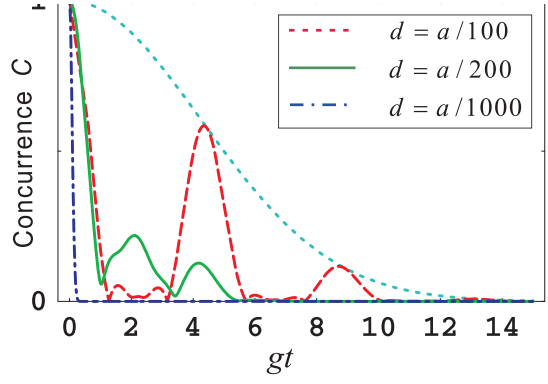}
\caption{(Color online) Time evolution of the concurrence $C$ for the case
when the initial internal state of the two atoms is an entangled state $%
\left\vert \protect\phi \left( 0\right) \right\rangle =\cos \protect\gamma %
\left\vert e_{1},g_{2}\right\rangle +\sin \protect\gamma \left\vert
g_{1},e_{2}\right\rangle $, with $\protect\gamma =\protect\pi /4$. The
dotted line is for $e^{-s}$ with $d=a/100.$}
\label{figdisentanglement2}
\end{figure}

The concurrence for the reduced density matrix $\hat{\rho}_{i}(t)$ in Eq. (%
\ref{partial density}) is given by $C\left( \hat{\rho}_{i}(t)\right)
=2\left\vert z\right\vert ,$ which is shown in Fig. \ref{figdisentanglement2}%
. One can see that there is a common factor $e^{-s(t)}$ with $s(t)=\hbar
^{2}k^{2}t^{2}/\left( 4d^{2}m_{0}^{2}\right) $, appearing in the expression
of $z$ above. Because of this factor $e^{-s(t)}$, the value of $\left\vert
z\right\vert $ decays to zero with time asymptotically. Here, Fig. \ref%
{figdisentanglement} and Fig. \ref{figdisentanglement2} show that the
oscillations of the entanglement are damped exponentially, but the decay of
the entanglement occurs faster when the width $d$ of the initial Gaussian
function becomes smaller, and the entanglement decays to almost zero in a
finite time. So the back-action of the emitted photons may induce an
entanglement between the two-atom internal states and the two-atom spatial
motion states, and this correlation destroys the coherence of the internal
states of the two atoms. The smaller the width of the wave packets, the
larger the uncertainty of the atomic spatial momentum, which results in a
faster destruction of the atomic internal-state entanglement.

\section{Conclusion}

We have considered a system of two atoms interacting with a single-mode
cavity field. Since the spatial degrees of freedom of the atoms are
considered, the Hamiltonian of the whole system becomes complicated. To
solve the Sch\"{o}rdinger equation, we have introduced two unitary
transformations (involving the coupling of the internal levels with the
spacial degrees of the atoms), presented an effective Hamiltonian of the
system, and given the analytic solutions to the eigenequation of the
effective Hamiltonian. Based on these, we have investigated the decoherence
dynamics for the relative position of the two atoms, and presented a
decoherence factor for a general cavity-field state and an arbitrary atomic
spatial state, which might be useful for future related works.

Under the assumption that the atomic relative position is in a superposition
of two Gaussian wave packets, we have demonstrated the spatial decoherence
of the two atoms' relative position, for a cavity field initially in a
coherent state. Our results show that the spatial decoherence of the two
atoms depends strongly on the relative position of the two atoms.
Interestingly, we found that when the relative position of the two atoms is
an integral multiple of the wavelength of the cavity field, the spatial
coherence of the relative position of the two atoms is not destroyed by the
photon recoil; However, when the relative position of the two atoms is not
an integral multiple of the wavelength of the cavity field, spatial
decoherence of the relative position of the two atoms happens.

Furthermore, we have studied the entanglement dynamics of the internal
states of the two atoms interacting with a single-mode cavity field. Our
results show that the entanglement, measured by the concurrence, decays to
almost zero in a finite time. Thus, the back-action of the photons emitted
from the two atoms may be a fundamental process destroying the entanglement
of atoms.

From this work, it can be concluded that there exists a phenomenon that the
moving qubits (e.g., atoms) placed in a cavity, may suffer from the spatial
decoherence. Therefore, it is important to overcome the influence of the
spatial motion of qubits on their entanglement.

\section{Acknowledgments}

We are grateful to C.P. Sun, A. Miranowicz, S.B. Zheng, X. G. Wang and P.
Zhang for very useful comments and help. FN acknowledges partial support
from the Laboratory of Physical Sciences, National Security Agency, Army
Research Office, National Science Foundation grant No. 0726909, JSPS-RFBR
contract No. 09-02-92114, Grant-in-Aid for Scientific Research (S), MEXT
Kakenhi on Quantum Cybernetics, and Funding Program for Innovative R\&D on
S\&T (FIRST). CPY acknowledges support from the National Natural Science
Foundation of China under Grant No. 11074062, the Natural Science Foundation
of Zhejiang Province under Grant No. Y6100098, and the funds from Hangzhou
Normal University. This work was also supported by the\textit{\ }China
Scholarship Council, and the Foundation of the Education Department of
Liaoning Province Grant No. 20060160.

\end{document}